\documentclass[a4paper,12pt]{article}
\pdfoutput=1
\usepackage[margin=1in]{geometry}
\usepackage{graphicx,xcolor}
\usepackage{mathtools,braket,blkarray}
\allowdisplaybreaks
\usepackage{amssymb,amsfonts,bbm,relsize}
\usepackage[width=0.9\textwidth,footnotesize]{caption}
\usepackage[font=scriptsize]{subcaption}
\usepackage{cite,hyperref,url}
\usepackage{verbatim}
\hypersetup{pageanchor=false}
\usepackage{booktabs,multirow,makecell}
\usepackage[utf8]{inputenc}

\newcommand{\diag}{\mathrm{diag}}

\begin{document}

\begin{titlepage}
\begin{center}
{\bf\Large   Gauge-flavon unification   } \\[12mm]
Alfredo Aranda$^{\ddag}$%
\footnote{E-mail: \texttt{fefo@ucol.mx}},
Francisco~J.~de~Anda$^{\dagger}$%
\footnote{E-mail: \texttt{fran@tepaits.mx}},
Stephen~F.~King$^{\star}$%
\footnote{E-mail: \texttt{king@soton.ac.uk}},
\\[-2mm]

\end{center}
\vspace*{0.50cm}
\centerline{$^{\ddag}$ \it
Facultad de Ciencias-CUICBAS, Universidad de Colima, C.P.28045, Colima, M\'exico 01000, M\'exico}
\centerline{Dual CP Institute of High Energy Physics, C.P. 28045, Colima, M\'exico}
\vspace*{0.2cm}
\centerline{$^{\dagger}$ \it
Tepatitl{\'a}n's Institute for Theoretical Studies, C.P. 47600, Jalisco, M{\'e}xico}
\vspace*{.20cm}
\centerline{$^{\star}$ \it
School of Physics and Astronomy, University of Southampton,}
\centerline{\it
SO17 1BJ Southampton, United Kingdom }
\vspace*{1.20cm}

\begin{abstract}
{\noindent
In this paper we propose the idea that flavons can emerge from extra dimensional gauge fields, 
referred to as gauge-flavon unification (GFU) analogous to gauge-Higgs unification (GHU).
We assume that there is a gauged family symmetry in extra dimensions and that the flavons are the extra dimensional components of the gauge field. This provides a simple mechanism to align the VEVs of the flavons through a combination of Wilson lines and orbifold symmetry breaking. We present some simple 5d examples of GFU based on $SO(3)$ and $SU(4)$ gauged family symmetry,
the latter case yielding $SU(3)\times U(1)$ gauged family symmetry in 4d, broken 
by triplet and antitriplet flavons, with effective couplings to fermions.
We also present a general formalism for Wilson lines and orbifolds, in any number of dimensions, including non-commutative aspects Wilson lines, which may be useful for aligning additional flavons as required for realistic models.  
}
\end{abstract}
\end{titlepage}

\section{Introduction}

Symmetry breaking plays an important role in particle physics. In the Standard Model (SM), the scalar Higgs field, is charged under the local gauged $SU(2)\times U(1)$ symmetry and acquires a vacuum expectation value (VEV) order parameter. 
In extra dimensions, gauge symmetries may be broken by the presence of background gauge fields on non-simply connected manifolds, a phenomenon known as the Hosotani mechanism~\cite{Hosotani:1983xw,Hosotani:1983vn}, or Wilson-line symmetry breaking~\cite{Candelas:1985en}. It is based on the Aharonov-Bohm observation that in the presence of a non-contractible loop, a gauge field can not always be gauged away and leads to shifts of momenta with observable effects. This mechanism is an extra dimensional version of spontaneous symmetry breaking (SSB), in which the extra dimensional components of gauge fields can effectively develop VEVs.

The idea of Wilson lines has been applied to the Higgs mechanism of the SM, in an approach known as 
Gauge-Higgs Unification (GHU) \cite{Sakamura:2007qz,Medina:2007hz,Hall:2001zb,Gogoladze:2003bb,Hosotani:2006qp,Scrucca:2003ut,Antoniadis:2001cv,Panico:2005dh,Adachi:2018mby}. This assumes extra dimensions with the Higgs boson coming from the extra dimensional part of the gauge fields. However this idea is severely challenged since it relates the compactification scale with the electroweak scale, which is tightly constrained. Nevertheless, it is interesting to consider a similar mechanism as applied to other areas of physics 
beyond the SM, in particular theories of flavour.

It is well appreciated that the SM does not address the flavour problem, namely the origin of the three fermion families and their pattern of masses and mixing angles. Some theories of flavour beyond the SM 
postulate a spontaneously broken family symmetry under which the three families form a triplet.
In such models, the spontaneous symmetry breaking
(SSB) of the family symmetry is achieved via the introduction of new triplet scalar fields called flavons that obtain VEVs which are aligned in various directions via an alignment potential (for a review see e.g. \cite{King:2013eh}).
Since such theories are rather involved, 
it is therefore natural to ask if extra dimensions can play a role in breaking the family symmetry, analogous to the idea of GHU discussed above.

In this paper we propose the idea that flavons can emerge from extra dimensional gauge fields, 
which we call gauge-flavon unification (GFU) in analogy to GHU.
GFU assumes that there is some gauged continuous non-Abelian family symmetry in extra dimensions, where the flavons 
emerge as the extra dimensional components of the gauge fields, after compactification. This identifies the compactification scale with the flavour breaking scale which is much much larger than the electroweak scale, making such an idea viable in principle. The extra dimensions allow simple mechanisms to align the VEVs of the flavons through a combination of Wilson lines and orbifolds, without the need for complicated alignment potentials. To illustrate the GHU mechanism, 
we present a simple example based on $SO(3)$ gauged family symmetry in 5d. This 
is the simplest choice since the adjoint representation is a real triplet representation, which is what we want for flavons.
We also discuss a $SU(4)$ gauged family symmetry in 5d which is also a natural choice since it can be broken through orbifolding to $SU(3)\times U(1)$ and the broken generators are complex triplets which may have effective couplings to fermions. 
Finally we present a general formalism for Wilson lines and orbifolds, in any number of dimensions, including non-commutative aspects Wilson lines, which may be useful for aligning additional flavons.  

The layout of the remainder of the paper is as follows. In section~\ref{Wilson5d} we review Wilson lines and orbifolds in 5d to show how the two mechanisms acting together may break the gauge group, reducing its rank, and yielding scalar fields as the extra dimensional components of the gauge fields, which may play the role of flavons. 
In section~\ref{so3} we apply these ideas to 
$SO(3)$ gauged family symmetry, in 5d, and show how triplet flavons with VEVs can emerge.
In section~\ref{su4} we discuss an $SU(4)$ gauge family symmetry model in 5d, 
and show how complex triplet and antitriplet flavons with VEVs can arise in this case, and show how they may have effective couplings to fermions in
a proto-model of quark and lepton flavour based on this mechanism.
In section~\ref{Wilson} we give a general definition of Wilson lines and orbifolds which may be extended to 
any number of dimensions, including non-commutative aspects Wilson lines, which may be useful for aligning additional flavons.  
Section~\ref{conclusion} concludes the paper.

\section{Wilson Lines and Orbifolds in 5d}
\label{Wilson5d}

Extra dimensions provide new mechanisms to break symmetries that are not available in 4d. In this section we review the Wilson line and orbifold mechanisms, to show how the two mechanisms acting together may break the gauge group, reducing its rank, and yielding scalar fields as the extra dimensional components of the gauge fields, which may play the role of flavons. 

\subsection{Wilson lines}
\label{sec:wl}

In four dimensional gauge theories one defines a gauge invariant Wilson loop as the trace of a path ordered (P) exponential of the gauge fields $A^a_{\mu}$ transported
along a closed line $C$,
\begin{equation}
W_C=P \exp i \oint_C T_aA^a_{\mu}dx^{\mu}
\label{W}
\end{equation}
where $T_a$ are the generators of the gauge group.
In the vacuum the associated field strength $F_{\mu \nu}$ can be zero. But this does not necessarily mean that the
gauge fields $A^a_{\mu}(x)$ in the exponent can be set equal to zero at all points along the path $C$ by a gauge transformation.
This is the Aharonov-Bohm observation.

Recall that in four dimensional gauge theories one can apply a general spacetime dependent gauge transformations on any field $\Psi(x)$, without affecting the physics
\begin{equation}
 T_aA^a_{\mu}\rightarrow \mathcal{U}(x) T_aA^a_{\mu}\mathcal{U}^{\dagger}(x) -i(\partial_{\mu} \mathcal{U}(x)) \mathcal{U}^\dagger(x), \ \ 
 \Psi(x)\rightarrow \mathcal{U}(x)\Psi(x), \ \ 
\mathcal{U}(x)=e^{i\alpha^a(x)T_a},
\label{G}
\end{equation}
where we absorb the gauge coupling constant into $A_\mu$. 

In the special case of a vacuum configuration,  i.e. if the field strength is equal to zero, it is possible to work in a gauge where the gauge fields $A^a_{\mu}$ are also 
equal to zero or a pure gauge. Here lies the usefulness of the Wilson line definition, where, for a pure gauge 
$ T_aA^a_{\mu}=-i\partial_{\mu} \mathcal{U}(x))\mathcal{U}^{\dagger}(x)$, it becomes
\begin{equation}
W_C=P \exp  \oint_C \partial_{\mu} \mathcal{U}(x))\mathcal{U}^{\dagger}(x) dx^{\mu},
\label{WC}
\end{equation}
where the the integral is identically zero for only smooth functions $\mathcal{U}(x)$. When these functions are ill defined at some point, the loop can't be contracted (due to the existence of a singular point) and it won't vanish. Therefore it can generate physical field strengths.

In extra dimensions the same effect can happen even when the gauge transformations are smooth. This happens if the extra dimensions are not a simply connected space. For simplicity let us assume a single compact extra dimension which is a circle $S^1$ with radius $R$.
In this case, the gauge transformation 
\begin{equation}
\mathcal{U}(y)=e^{i\alpha^a(y)T_a},
\end{equation}
which only depends on the extra dimension $y$, becomes physical. This is measured through the 
phase (c.f. Eq.~\ref{WC}),
\begin{equation}
U=P e^{\oint \partial_{5} \mathcal{U}(y))\mathcal{U}^{\dagger}(y)dy}=P e^{ \int_0^{2\pi R}\partial_{5} \mathcal{U}(y))\mathcal{U}^{\dagger}(y) dy}=e^{i\beta^a T_a},
\end{equation}
which arises from an integration through a non contractible loop and can't be trivialized \cite{Hall:2001tn}. In this case, the loop is non contractible because the space is not simply connected (a circle) even with smooth functions. 
Any field transported around this non contractible loop will pick up the phase $U$. However, going around the whole loop brings us back to the same point, and due to the continuity of the field function this phase acts as a boundary condition
\begin{equation}
\Psi(x,y)=U\Psi(x,y+2\pi R)=e^{i\beta^a T_a}\Psi(x,y+2\pi R),
\label{eq:psi2pir}
\end{equation}
from which we can obtain the explicit dependence of the extra coordinate $y$
\begin{equation}
\Psi(x,y)=e^{-i\beta^a yT_a/2\pi R}\tilde{\Psi}(x,y),\ \ \ \tilde{\Psi}(x,y)=\tilde{\Psi}(x,y+2\pi R),
\label{eq:solpsi}
\end{equation}
where $\tilde{\Psi}$ are periodic functions in $y$. If we move around a full circle $y\to y+2\pi R$, we recover Eq. \ref{eq:psi2pir}. 
We can apply a gauge transformation throughout
\begin{equation}
\Psi(x,y)\to e^{i\beta^a yT_a/2\pi R}\Psi(x,y)=\tilde{\Psi}(x,y),
\label{eq:U}
\end{equation}
so that all fields are periodic in the extra dimension. This would gauge away its effect from all the fields except the gauge fields that change as
\begin{equation}\begin{split}
A_m^bT_b&\to \tilde{A}_m^bT_b+(\partial_5 e^{i\beta^b yT_b/2\pi R}) e^{-i\beta^a yT_a/2\pi R}
\\ &= \tilde{A}_m^bT_b+\delta_m^5 \beta^bT_b/2\pi R,
\label{Wilsonvev}
\end{split}\end{equation}
where $m=0,1,2,3,5$. This adds a constant to the 5th component of the gauge field, generating an effective VEV and breaking the symmetry spontaneously. Note that the VEV can only happen in the 5th component so that 4d Lorentz symmetry is not broken. This is the Wilson line mechanism to break the gauge symmetry \cite{Candelas:1985en}. 

Note that the effective VEV is in the adjoint representation and it can't reduce the rank of the gauge group.

\subsection{Orbifolds}

In 5d the simplest orbifold is $S^1/Z_2$. The orbifolding happens through the identification
\begin{equation}
y\sim-y.
\end{equation}
This forces boundary conditions on the fields which have to comply with the identification, up to a phase

\begin{equation}
\Psi(x,y)=P\Psi(x,-y)=e^{i\alpha^aT_a}\Psi(x,-y),
\label{eq:orcon}
\end{equation}
where $P=e^{i\alpha^aT_a}$ is an arbitrary 
non-Abelian phase satisfying $P^2=1$, since it accompanies the parity transformation $\mathcal{P}_5$. The phase $P$ belongs to the gauge group and we assume that $\Psi$ is in the fundamental representation.

 This phase $P$ affects the gauge fields $A_m=A_m^a T_a$ in the adjoint representation as
 \begin{equation}
 A_\mu(x,y)=A_\mu(x,-y)^a\ PT_a P,\ \ \ A_5(x,y)=-A_5(x,-y)^a\ PT_a P,\end{equation}
 where $T_a$ are the generators of the gauge group. The 5th component receives an extra minus sign since it is part of a vector field and we are applying a parity transformation on the 5th coordinate. After compactification all fields are decomposed into an infinite tower of KK modes. Only the fields that have trivial boundary condition have a massless zero mode and survive at low energies. 

In the case where $P=1$, the unit matrix, after compactification, the 5d gauge vector is separated into the 4d gauge vector $A_\mu$ that has a positive eigenvalue and therefore a massless zero mode. The remaining component $A_5$ behaves like a 4d scalar and has a negative eigenvalue, therefore lacking a massless zero mode and does not survive at low energies.

In the case with general $P$, it must comply with $P^2=1$ from which we know that it only has eigenvalues $\pm 1$. This condition separates the  generators  $\{T_a\}=\{t_b,\tilde{t}_{\hat{b}}\}$ where
\begin{equation}
P t_b P= t_b, \ \ \ P \tilde{t}_{\hat{b}} P=-\tilde{t}_{\hat{b}}.
\label{eq:hatb}
\end{equation}
Therefore only the vector fields $A_\mu^b$ corresponding to parity even generators
have a zero mode and, at low energies, the gauge group is broken into the subgroup generated by $t_b$, i.e. the ones that commute with $P$. 
Although the vector fields $A_\mu^{\hat{b}}$ corresponding to parity odd generators are massive,
the 4d scalars $A_5^{\hat{b}}$ corresponding to the broken generators now have a zero mode, as we now discuss.

Since the orbifold operation $y=-y$ implies a parity transformation,  there are two sources of a negative eigenvalue for the fields. The first one coming from the Lorentz representation, separating the extra dimensional component ($m=5$) from the 4d ones ($m=\mu$). The second one comes from the gauge representation, separating the broken generators ($\tilde{t}_{\hat{b}}$) from the unbroken ones ($t_b$). Each contribution is shown separately for each component of the gauge field, in the table \ref{tab:1}. Only the fields with final 
combined $+1$ eigenvalue have zero mode.
\begin{table}[ht]
	\centering
	\begin{tabular}[t]{|c|cc|}
		\hline
		$A_m^b T_b$ & $T_b=t_b$ & $T_b=\tilde{t}_{\hat{b}}$ \\ 
		\hline
	$m=\mu $ & $(1)(1)$ & $(1)(-1)$ \\
	$m=5 $ & $(-1)(1)$ & $(-1)(-1)$ \\
		\hline
	\end{tabular}
	\caption{Eigenvalues of $A_m^b$ under $(P)(\mathcal{P}_5)$ respectively for each component. The components $A_\mu^b, A_5^{\hat{b}}$ have
	a combined $+1$ eigenvalue and hence have massless modes.}
	\label{tab:1}
\end{table}

Then we conclude that the gauge transformation $P$ breaks the gauge symmetry at low energies into the subgroup that commutes with it. Furthermore it allows the 4d scalars corresponding to the broken generators to have zero modes.

Note that the $P$ by itself breaks the gauge symmetry at low energies but it can't reduce the rank of the remaining gauge group. The effective 4d scalars with zero modes come from the  broken generators, and if they obtain a VEV, they would reduce the rank. However they come from the extra dimensional gauge vector component and they do not have a potential that gives them a VEV.
As we discuss in the next subsection, a mechanism capable of giving the 4d scalars with zero modes a VEV may be provided
by Wilson lines. In this way, the combination of Wilson lines and orbifolds may break the rank of the gauge group.

\subsection{Wilson lines and orbifolds}

The simplest setup to study orbifolds together with Wilson lines happens in 5d and the orbifold $S^1/Z_2$. This orbifold is defined by the conditions on the extra dimension
\begin{equation}\begin{split}
y&\sim y+2\pi R,\\
y&\sim -y,
\label{eq:y}
\end{split}\end{equation}
where the first equation defines compactification on a circle while the second one is the orbifolding identification. Actually, as we will see in 
a later section, this is the only consistent orbifolding identification in 5d.
Any field $\Psi(x,y)$ propagating through the 5th dimension would be subject to comply with the orbifold conditions from eq. \ref{eq:y}. In the case, where there is a gauge symmetry, the field must comply with them up to the arbitrary phases introduced earlier,
\begin{equation}\begin{split}
\Psi(x,y)&=U\Psi(x,y+2\pi R),\\
\Psi(x,y)&=P\Psi(x,-y),
\label{eq:pu}
\end{split}\end{equation}
where $U,P$ are equivalent to applying Eqs. \ref{eq:psi2pir},\ref{eq:orcon} together.
They must satisfy the consistency conditions
\begin{equation}
P^2=1,\ \ \ PUPU=1,
\label{eq:pupu}
\end{equation}
coming from the fact that they are related to a translation and a reflection as discussed below
(see also \cite{deAnda:2019anb,Hebecker:2003jt,Hebecker:2001jb}).

The Wilson line mechanism and orbifolding break the 5d Poincar\'e symmetry but respect the 4d part. The non trivial $U$ breaks translation symmetry under the 5th coordinate while a non trivial $P$ breaks the corresponding parity. This is the most general breaking for 5d. In higher dimensions, rotations can also be broken as discussed in section~\ref{Wilson}.

This orbifold has two fixed branes (at $y=0,\pi R$) and the gauge tranformations $U,P$ define its boundary conditions which act on the fields as
\begin{equation}
\Psi(x,y)=P_0\Psi(x,-y),\ \ \ \Psi(x,y+\pi R)=P_{\pi R} \Psi(x,-y+\pi R).
\end{equation}
where $P_{0,\pi R}$ are the corresponding boundary conditions at $0,\pi R$. These are uniquely fixed by the gauge transformations which act as boundary conditions from Eq. \ref{eq:pu} to be
\begin{equation}
P_0=P,\ \ \ P_{\pi R}= PU,
\end{equation}
where the second equality comes from from applying the reflection and then the translation
\begin{equation}
\Psi(x,y+\pi R)=P\Psi(x,-y-\pi R)=PU\Psi(x,-y-\pi R+2\pi R)=PU\Psi(x,-y+\pi R).
\end{equation}
From this equation we can see that $(PU)^2=1$ implying the condition on Eq. \ref{eq:pupu}.

The orbifold condition $P$ breaks the gauge symmetry by modding out zero modes. The Wilson line breaking through $U$ induces an effective VEV in the scalar coming from the extra dimension.
However the orbifold consistency condition from eq. \ref{eq:pupu} restricts the possible VEVs 
\begin{equation}
1=PUPU=Pe^{i\beta^aT_a}Pe^{i\beta^aT_a}=e^{i\beta^a PT_aP}e^{i\beta^aT_a},
\end{equation} 
where the last equality comes from the fact that $P^2=1$ and it can be inserted anywhere in the expansion of the exponential. This equation is only satisfied if the phase is aligned only with the $P$ broken generators $t_{\hat{a}}$ from Eq. \ref{eq:hatb} since
\footnote{A special case to solve $1 = P U P U$ with $P^2 = 1$ is given by choosing a $U$ with $P U = U P$. Then $U^2 = 1$ must hold. A similar mechanism of rank reduction was also discussed in~\cite{Hebecker:2003jt}.} 
\begin{equation}
e^{i\beta^{\hat{a}} Pt_{\hat{a}}P}e^{i\beta^{\hat{a}}t_{\hat{a}}}=e^{-i\beta^{\hat{a}}t_{\hat{a}}}e^{i\beta^{\hat{a}}t_{\hat{a}}}=1.
\end{equation}
Therefore the effective VEV in the scalar coming from the extra dimension from Eq.\ref{Wilsonvev} is
\begin{equation}
\braket{A_5^{\hat{b}}}=\frac{\beta^{\hat{b}}}{2\pi R}.
\end{equation}
This means that only the scalars coming from the fifth component of the gauge field that have a zero mode after orbifolding, can obtain an effective through the Wilson line. In this work we identify these scalars with the flavons
\begin{equation}
A_5^{\hat{b}}\sim \phi,
\end{equation}
whose VEV and its non-Abelian vacuum alignment is fixed through a choice of Wilson lines. 
To be precise, before orbifolding (and before Wilson lines) the gauge boson $A_5$ is in the adjoint of the higher-dimensional gauge group $G$. After orbifolding, $A_5$ is from the ``Coset G/H'' where $H$ is the unbroken subgroup of $G$. E.g. If one starts with $SU(5)$ and breaks it to $SU(4) \times U(1)$ by orbifolding, $A_5$ has massless modes only in the $4 + \bar{4}$ of $SU(4)$.

 By integrating out the matter content of a model, one can obtain an effective potential for the flavons
\cite{Hosotani:1983xw,Hosotani:1983vn,Hosotani:2004wv,Hosotani:2004ka,Haba:2004qf,Haba:2002py}. This potential would have the effect of driving and aligning the flavon VEVs. The specifics of the potential are model dependent, however they don't affect our general discussion. The examples given below are consistent even after considering the effective potential since any alignment can be rotated into the mentioned one, due to the continuous nature of the flavour symmetry.

To summarise the mechanism we have been developing,
the orbifolding process selects which zero modes survive after compactification by imposing boundary conditions. This effectively breaks the symmetry at low energies and may allow scalars with massless modes. The Wilson Line mechanism generates an effective VEV in a scalar coming from the extra dimensional part of the gauge fields. This scalar VEV must also comply with the boundary conditions coming from orbifolding. Therefore when applying both mechanisms of breaking the symmetry, the VEVs coming from the Wilson Line must lie in the rank-breaking direction of the scalar with the zero mode, thereby breaking the rank of the gauge group \footnote{There can exist rank preserving VEVs in the case that the group has generators that can  satisfy $e^{i2\beta^aT_a}=0$ for nonzero $\beta^a$.}.
We emphasise that the orbifold breaking alone does not reduce the rank of the gauge group. With both orbifold and Wilson line breaking, 
the VEV must be aligned along the rank breaking direction, thereby reducing the rank of the remaining gauge group.

\section{$SO(3)$ Gauge-Flavon Unification}
\label{so3}


As a first example of the preceding formalism,
consider the gauge group $SO(3)$ in 5d, with an orbifold $S^1/Z_2$,
where the extra dimensional part of the gauge fields are scalars which transform as triplets under $SO(3)$. If this is chosen to be a gauged family symmetry, these fields are identified as flavons. 

The VEVs of such flavons can be achieved via the Wilson line mechanism as described in Section \ref{Wilson5d}. We would have as many flavons as extra dimensions (in this case one) and their alignment is fixed by the choice of the Wilson line phase.

In the presence of both Wilson lines and orbifold boundary conditions, we need to consider the
generators of the $SO(3)$ algebra which are
\begin{equation}
T_{1}=\left(\begin{array}{ccc}0 & 0 & 0 \\ 0 & 0 & -1 \\ 0 & 1 & 0\end{array} \right),\ \ T_{2}=\left(\begin{array}{ccc}0 & 0 & -1 \\ 0 & 0 & 0 \\ 1 & 0 & 0\end{array} \right),\ \ T_{3}=\left(\begin{array}{ccc}0 & -1 & 0 \\ 1 & 0 & 0 \\ 0 & 0 & 0\end{array} \right),
\end{equation}
Any element of the algebra of $SO(3)$, and therefore any phase from a Wilson line as in Eq. \ref{eq:psi2pir}, can be written as
\begin{equation}
U=e^{i\beta^aT_a},  \ \ \ \ 
T_a\beta^a=\left(\begin{array}{ccc}
0&\beta^3&\beta^2\\
-\beta^3&0&\beta^1\\
-\beta^2&-\beta^1&0
\end{array}\right),
\label{eq:uso3}
\end{equation}
with real $\beta^a$. This can be rewritten in the usual flavon notation as a triplet column vector
\begin{equation}
\braket{\phi}=\frac{1}{2\pi R}\left(\begin{array}{c}\beta^1\\ \beta^2 \\ \beta^3\end{array}\right),
\label{eq:uso31}
\end{equation}
so that we can obtain any desired real alignment by choosing the phase.\footnote{We shall assume that the 
Wilson lines are continuous, i.e. the phases in Eqs.\ref{eq:uso3},\ref{eq:uso31} are not quantised, which may not always be the case~\cite{Forste:2005rs}.}

If we only have the Wilson line, without any $SO(3)$ orbifold breaking condition, it can always be rotated to the $(0,0,1)^T$ direction, which breaks $SO(3)\to U(1)$,
preserving the rank.

On the other hand if we consider orbifold breaking without Wilson lines, the parity $P$ boundary condition can always be rotated to be
\begin{equation}
P=e^{i\pi T_{3}}=\left(\begin{array}{ccc}-1&0&0 \\ 0 & -1 & 0\\ 0 & 0 &1 \end{array}\right),
\label{eq:pso3}
\end{equation}
 which breaks $SO(3)\to U(1)$ (and also does not reduce rank). 

However, as discussed in the previous section, the combination of Wilson lines and orbifold boundary conditions can reduce the rank of the remaining symmetry group. The boundary condition $P$ leaves an $U(1)$ invariant and the Wilson Line can give a VEV in that remaining $U(1)$.

Imposing the $U$ from Eq. \ref{eq:uso3} together with the $P$ from Eq. \ref{eq:pso3}, they must comply with the consistency condition from  Eq. \ref{eq:pupu}
\begin{equation}\begin{split}
PUPU&=1\\
&=(Pe^{i\beta^aT_a}P)e^{i\beta^aT_a}=e^{i\beta^aPT_aP}e^{i\beta^aT_a}\\
&=e^{i(-\beta^1T_1-\beta^2T_2+\beta^3T_3)}e^{i(\beta^1T_1+\beta^2T_2+\beta^3T_3)},
\end{split}\end{equation}
 which has 2 nontrivial solutions. 
The first one is where $\beta^3=0$ and general real $\beta^{1,2}$.
This is equivalent to a flavon alignment
\begin{equation}
\braket{\phi}=\frac{1}{2\pi R}\left(\begin{array}{c} \beta^1\\ \beta^2 \\ 0
\end{array}\right),
\end{equation}
which breaks the remaining $U(1)$.
However in this example there is also a rank preserving solution.
\footnote{Note that there is a second rank preserving 
solution when $\beta^1=\beta^2=0$ and $\beta^3=\pi$ which is equivalent to the flavon alignment
\begin{equation}
\braket{\phi}=\frac{1}{2 R}\left(\begin{array}{c} 0\\ 0 \\ 1
\end{array}\right),
\end{equation}
which preserves the remaining $U(1)$ and has no freedom on the absolute value.
Note that this is consistent with the fact that any field, must be an eigenstate of the P operator. Therefore the allowed VEV alignments are the eigenvectors of the $\diag(-1,-1,1)$ matrix.
}

In the $SO(3)$ gauge theory with 6 dimensions where the orbifold is $T^2/\mathbb{Z}_2$, there would be
two flavons obtained. There would also be two possible Wilson lines 
associated with each extra dimensions 
\begin{equation}
U_5=e^{i\beta^a_5T_a},\ \ \ U_6=e^{i\beta^a_6T_a}.
\end{equation}
Coming from translations of independent circles, these phases must commute $[U_5,U_6]=0$, and this fixes
\begin{equation}
\beta^a_5\propto\beta^a_6,
\end{equation}
which means that the two flavon VEVs must be aligned with each other so nothing is gained by having two flavons with identically 
aligned VEVs.
However, we can override this condition assuming an $SO(3)$ gauge theory, where there is a non-commutative twist in the fibre bundle as described in Sec. \ref{sec:noncom}, leading to non-aligned VEVs in that case. However for the moment we shall continue to consider the usual commutative case, and go on to consider another example.

\section{$SU(4)$ Gauge-Flavon Unification}
\label{su4}

\subsection{Flavon sector}
As a second example,
let us assume  that we have a $SU(4)$ gauge family symmetry model in $5$ dimensions
with an orbifold $S^1/Z_2$.
Since this group allows complex representations, it will allow two non-aligned flavons
in conjugate representations.
In this case, let us assume that the extra dimensions are orbifolded with a twist on the ED parity
\begin{equation}
P_M\sim\diag(1,1,1,-1),
\end{equation}
that breaks $SU(4)\to SU(3)\times U(1)$.
The generators $T_a$ with $a=1,...,15$ decompose as
\begin{equation}\begin{split}
T_a&=\lambda_a,\ \ \ {\rm for}\ \ \ a=1,...,8,\ \ \ T_{15}=\diag(-1/3,-1/3,-1/3,1)\\
T_9&=
\left(\begin{array}{cccc} 0&0&0&1\\0&0&0&0\\0&0&0&0\\1&0&0&0
\end{array}\right),\ \ \ T_{11}=
\left(\begin{array}{cccc} 0&0&0&0\\0&0&0&1\\0&0&0&0\\0&1&0&0
\end{array}\right),\ \ \ T_{13}=
\left(\begin{array}{cccc} 0&0&0&0\\0&0&0&0\\0&0&0&1\\0&0&1&0
\end{array}\right),\\
T_{10}&=
\left(\begin{array}{cccc} 0&0&0&-i\\0&0&0&0\\0&0&0&0\\i&0&0&0
\end{array}\right),\ \ \ T_{12}=
\left(\begin{array}{cccc} 0&0&0&0\\0&0&0&-i\\0&0&0&0\\0&i&0&0
\end{array}\right),\ \ \ T_{14}=
\left(\begin{array}{cccc} 0&0&0&0\\0&0&0&0\\0&0&0&-i\\0&0&i&0
\end{array}\right),
\end{split}\end{equation}
where the first line contains the Gell-Mann matrices $\lambda_a$, the generators of the unbroken $SU(3)$ and $T_{15}$ the generator of the unbroken $U(1)$. The Wilson line phase can be written $U=e^{i\beta^aT_a}$ as in Eq. \ref{eq:uso3} and it must comply with
the consistency condition from  Eq. \ref{eq:pupu}
\begin{equation}
1=P_MUP_MU=P_M e^{i\beta^aT_a}P_Me^{i\beta^aT_a}=e^{i\beta^a(P_MT_aP_M)}e^{i\beta^aT_a}.
\end{equation}
It can easily be seen that
\begin{equation}\begin{split}
P_M T_a P_M=T_a,\ \ \ &{\rm for} \ \ \ a=1,...,8,15,\\
P_M T_a P_M=-T_a,\ \ \ &{\rm for} \ \ \ a=9,...,14,
\end{split}\end{equation}
so that $U$ can only have the generators $T_a$ with $a=9,...,14$.
This way the gauge fields are broken by the boundary condition $P_M$  into 
\begin{equation}
15\to 8_0+1_0+3_{-4/3}+\bar{3}_{4/3},
\end{equation}
where the subscript indicate the $U(1)$ charge. The Wilson line restricted by $P_M$ can only give VEVs to the triplets
and antitriplets.

After orbifolding we have the effective scalars corresponding to the extra dimensional part of the broken gauge fields behaving like flavons. The gauge fields decompose as
\begin{equation}
A_m=\left(\begin{array}{cc} G_\mu-Z'_\mu/3 & \phi\\
\bar{\phi}& Z'_\mu
\end{array}\right), 
\end{equation}
where the flavons in the $3,\overline{3}$ conjugate representations of $SU(3)$ are
\begin{equation}
\phi=\frac{1}{2}\left(\begin{array}{c}A_5^{9}+iA_5^{10} \\ A_5^{11}+iA_5^{12} \\A_5^{13}+iA_5^{14}\end{array}\right),\ \ \ \ \bar{\phi}=\frac{1}{2}\left(A_5^{9}-iA_5^{10}, A_5^{11}-iA_5^{12},A_5^{13}-iA_5^{14}\right),
\end{equation}
which can obtain general VEVs by fixing the Wilson line phase. The flavons, coming from the gauge fields, are real scalars  arranged into the complex triplet and antitriplet representations as shown above which limits the possible VEVs. 
We note that there is only one independent complex scalar triplet and the other one is its complex conjugate. A general VEV in this triplet can always be rotated into the form
\begin{equation}
\braket{\phi}\sim \left(\begin{array}{c} 0\\0\\a\end{array}\right),\ \ \ \braket{\bar{\phi}}\sim (0,0,a),
\label{eq:su4abc}
\end{equation}
with arbitrary real $a$. In general this alignment breaks the $SU(3)\times U(1)\to SU(2)\times U(1)$ reducing the rank by 1. 

The orbifold condition breaks $SU(4)$ by only allowing zero modes to the gauge vectors of $SU(3)\times U(1)$. The flavons further break this symmetry by obtaining a VEV. Usually it is necessary for the flavour symmetry to be completely broken to generate the observed fermion masses and mixings. With this remaining symmetry we would need at least 3 flavons to achieve it, which would come from at least 3 extra dimensions.

Having more than one extra dimension will yield extra flavon pairs in conjugate representations. The commutation requirement between Wilson lines in this $SU(4)$ theory constrains the possible alignments of the second pair to be aligned as the first one. If one wants different flavon alignments 
from those above, one must consider the non commutative Fibre Bundle construction described in Sec. \ref{sec:noncom}.
However it is possible to envisage a proto-model just with the two flavons above, along the following lines.

\subsection{Towards a model with fermions}

Let us assume all left-handed SM fermions $f= L,Q$ and $f^c=e^c,\nu^c,u^c,d^c$ transform as $4$ dimensional 
representations of $SU(4)$ and propagate in the bulk.  The boundary condition breaks the multiplet $4\to 3_{-1/3}+1_1$ and only the triplet has a zero mode. The fermions modes would be denoted as $(f,f^c)_{3,1}^{n}$, where the subscript labels the $SU(3)$ triplet or singlet component and the superscript is $0$ for the zero mode and $KK$ otherwise. The low energy fermions are always the zero mode triplets $f,f^c=(f,f^c)^0_3$ so that their sub-indices and super-indices can be suppressed without confusion.

The Higgs doublet is fixed on a brane, which only has $SU(3)\times U(1)$ flavour symmetry, and is chosen to be an $SU(3)$ 
flavour singlet with a $U(1)$ charge of $-2$. There is no  ordinary Yukawa term for the zero modes, so that all of them are higher order. However the Higgs does have a renormalizable coupling to $SU(3)$ singlet fermions.

There are two effective flavon triplets for each extra dimension $\phi_i,\bar\phi_i$. After integrating the KK modes we can obtain the effective Yukawa terms
\begin{equation}
\mathcal{L}_Y\sim Tr\Big[f f^c H  \left(\bar{\phi}\bar{\phi}+\bar\phi\phi^\dagger+\phi^\dagger\bar{\phi}+\phi^\dagger\phi^\dagger\right)\Big],
\label{effyuk}
\end{equation}
where all the $f, f^c$ are the $SU(3)$ triplet zero modes and the corresponding subscripts and superscripts are suppressed. If there is more than one extra dimension, there would be these terms for each flavon as well as all possible mixings.

\begin{figure}[h!]
		\centering
		\includegraphics[scale=0.4]{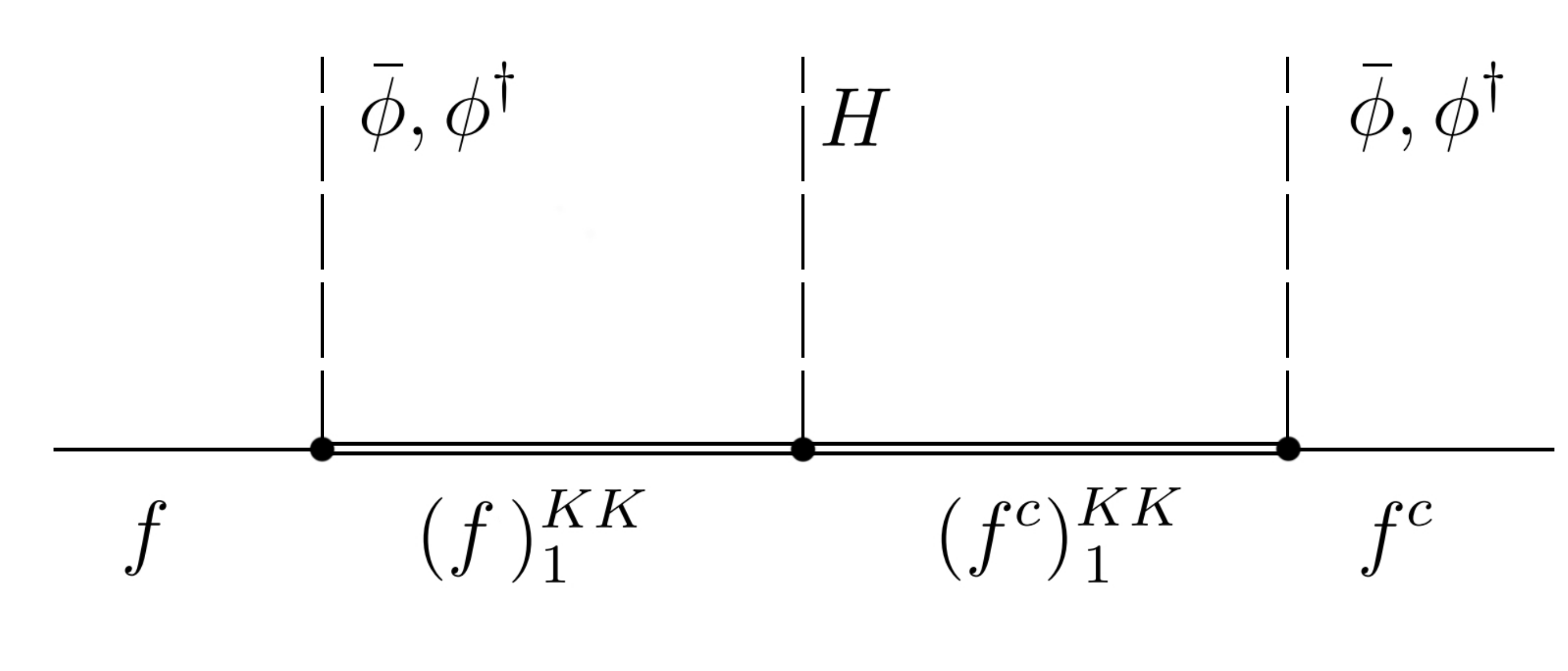}
	\caption{Diagram that generates the effective Yukawa terms mediated by KK modes.}
	\label{fig:kk}
\end{figure}

The terms in Eq.~\ref{effyuk} come from diagrams as in figure \ref{fig:kk}. The flavon insertions can be either $\bar{\phi}$ or $\phi^\dagger$ of which there are 4 possibilities and generate the 4 different terms. The KK modes are always Dirac fermions, have masses of the order of the compactification scale and do the chirality flip by themselves. The KK modes mediate all higher order terms and their couplings to the zero modes and gauge fields are fixed, which fixes completely the Yukawa couplings

Although the heavy KK modes mediate the effective Yukawa couplings, such diagrams are not very 
suppressed since the KK mode mass and the Wilson line VEVs $\braket{\phi,\bar{\phi}}$ are both the same order, namely the compactification scale. KK mode mediated higher order terms are possible but all involve insertions of $\phi\bar\phi$ contractions which do not change the mass matrix structure. This way the first order terms completely determine the structure of the mass matrices. 

A model along these lines would have an effective $SU(3)$ gauged flavour symmetry where all fermions are triplets.
Flavons in the triplet and antitriplet representation are generated automatically with aligned VEVs, resulting from
the extra dimensional components of the underlying $SU(4)$ gauged flavour symmetry. 
No flavons need to be added by hand, nor is any alignment potential required.
However a realistic model would require additional flavons with independent vacuum alignments. This requires some further discussion
about Wilson lines and orbifolds,
and this is the subject of the next section.

\section{Wilson Lines and Orbifolds Revisited}
\label{Wilson}

We now present a general definition of the Wilson Lines together with the orbifold conditions where we assume a spacetime with $M$ dimensions, where $m=0,...,M$.
Such a formalism may be necessary for a realistic model where several flavons with independent vacuum alignments
are required.

\subsection{Wilson lines and fibre bundles }
In compact 5d periodic spaces the Wilson line is defined by the phase in Eq. \ref{eq:psi2pir}. This can be rewritten with the translation operator in the 5th dimension
\begin{equation}
\Psi(x,y+2\pi R)=e^{2\pi R \partial_5}\Psi(x)=e^{i2\pi R P_5}\Psi(x)=e^{i\beta^a T_a}\Psi(x),
\end{equation}
where $P_5$ is the momentum operator which is the generator of translations. We can see that defining the Wilson line gauge transformation is equivalent to changing the momentum operator to include an element of the gauge algebra
\begin{equation}
P_5\to P_5-\beta^a T_a /2\pi R.
\end{equation}

The Wilson line mechanism can be implemented in any number of dimensions by generalizing the momentum operator to include an element of the gauge algebra
\begin{equation}
P_m\to P_m-\beta^a_m T_a/2\pi R_m,
\end{equation}
which would generate a VEV in the same way described in section \ref{sec:wl} for the extra dimensional components of the gauge fields
\begin{equation}
\braket{A_m^a}=\beta^a_m/2\pi R_m,
\end{equation}
generalizing Eq. \ref{Wilsonvev} for more extra dimensions.

In gauge theories, the fields do  not live only in the Minkowski manifold but in a fibre bundle whose base is the Minkowski spacetime $\mathcal{M}^M$ and the fibre is a Lie group $G$. A gauge theory is built in such way that all physical observables are blind to the choice of a section (local gauge invariance).
This means that the fibre bundle is not a trivial product $\mathcal{M}^M\times G $ since it has a twist defined by $\beta^a_m$. 
This induces a preferred section of the fiber bundle, which is similar (and in the case where $\alpha$ a linear function, equivalent) to a Spontaneous Symmetry Breaking (SSB). However this SSB happens in the gauge group as well as in the Poincar\'e group, breaking Lorentz symmetry.
4d Lorentzian symmetry must be unbroken, but in ED, the extra dimensional part of the Poincar\'e group disappear after compactification and they can be broken.

To summarize, the Wilson line mechanism is equivalent to choosing a twist in the fibre bundle so that the translation operator also involves a gauge transformation, leaving a preferred section and inducing SSB. Choosing a specific twist means to choose the VEV alignment.

\subsection{Wilson lines and the Poincar\'e group}

We can generalize any of the transformations generated by the extra dimensional Poincar\'e group, not just translations. To generalize the rotation operator we need at least 2 extra dimensions.

Wilson lines are generalizations of the extra dimensional translation operators which generate an effective VEV and spontaneous symmetry breaking. Orbifold identifications are related to generalizations of the extra dimensional parity operators $\mathcal{P}_m$ and extra dimensional rotations which are defined by the Lorentz generators $M_{mn}$. These break the symmetry at low energies by only leaving zero modes to the fields that commute with the chosen gauge transformation. The orbifold identification must be a discrete operation from the group $SO(M-4)\times \mathbb{Z}_2^{M-4}$, leaving the 4d Poincar\'e group unbroken.

These generalizations can be applied simultaneously to the extra dimensional part of the Poincar\'e group $SO(M,1)\ltimes T^M$ when complying with its algebra
\begin{equation}\begin{split}[P_m, P_n] &= 0,\\
\frac{ 1 }{ i }~[M_{mn}, P_r] &= \eta_{mr} P_n - \eta_{nr} P_m,\\
\frac{ 1 }{ i }~[M_{mn}, M_{rs}] &= \eta_{mr} M_{ns} - \eta_{ms} M_{nr} - \eta_{nr} M_{ms} + \eta_{ns} M_{mr},
\label{eq:poin}
\end{split}\end{equation}
which is generalized to $O(M,1)\ltimes T^M$ by including the parity operators
\begin{equation}
\begin{split}
(\mathcal{P}^m)^2&=1,\\
[\mathcal{P}^m,P_n]&=-2\delta^m_n\ P_n\mathcal{P}_m,\\
[\mathcal{P}^m,M_{rs}]&=-2(\delta^m_r+\delta^m_s)\ M_{rs}\mathcal{P}_m,
\end{split}
\end{equation}
where this limits the possible VEVs that can be generated through Wilson lines in a specific orbifold. 

We conclude that using the Wilson line mechanism together with orbifolding in any number of dimensions is equivalent to generalizing the momentum, Lorentz and parity operators to include an element of the algebra of the gauge group $G$ generated by $T_a$
\begin{equation}
\begin{split}
\mathcal{P}^m &\to e^{i\alpha^{ma}T_a} \mathcal{P}^m,\\
P_m &\to P_m-\beta^a_m T_a/2\pi R_m,\\
M_{mn}&\to M_{mn}-\gamma^a_{mn}T_a,
\end{split}
\end{equation}
where all indices must be $m>3$ to preserve 4d Poincar\'e symmetry. Since the canonical Poincar\'e generators obviously comply with the algebra, the chosen twist in $(O(M,1)\ltimes T^M)\times G$ must comply with
\begin{equation}
\begin{split}
[T_a, T_b]&=f_{ab}^{\ \ c}T_c,\\
\beta^a_m \beta^b_nf_{ab}^{\ \ c} &= 0,\\
\frac{ 1 }{ i } \gamma_{mn}^a\beta_r^bf_{ab}^{\ \ c} &= \eta_{mr} \beta_n^c - \eta_{nr} \beta_m^c,\\
\frac{ 1 }{ i }~\gamma_{mn}^a\gamma_{rs}^b f_{ab}^{\ \ c} &= \eta_{mr} \gamma_{ns}^c- \eta_{ms} \gamma_{nr}^c - \eta_{nr} \gamma_{ms}^c + \eta_{ns} \gamma_{mr}^c,\\
(e^{i\alpha^{ma}T_a})^2&=1,\\
e^{i\alpha^{ma}T_a}\beta_n^cT_c&=-\delta_n^m \beta_n^c T_ce^{i\alpha^{ma}T_a},\\
e^{i\alpha^{ma}T_a}\gamma_{rs}^c T_c&=-(\delta^m_r+\delta^m_s)\ \gamma_{rs}^c T_ce^{i\alpha^{ma}T_a},
\label{eq:poinga}
\end{split}
\end{equation}
where $f_{ab}^{\ \ c}$ are the structural constants of the gauge algebra. The constraints regarding parity are simplified
In the 5d case, these constraints reduce to Eq. \ref{eq:pupu}.

Even though commuting Wilson lines by themselves do not reduce rank, together with the generalized parity and rotation transformations, rank reduction may happen \cite{Forste:2005rs}. 

\subsection{Non-commuting Wilson lines}
\label{sec:noncom}

In this subsection, we focus in the case with 2 extra spatial dimensions.
In a compact periodic space any field would have the independent boundary conditions
\begin{equation}
\begin{split}
\Psi(x,y_5+2\pi R_5,y_6)&=e^{i\beta_5^a T_a}\Psi(x,y_5,y_6),\\
\Psi(x,y_5,y_6+2\pi R_6)&=e^{i\beta_6^a T_a}\Psi(x,y_5,y_6),
\label{eq:psi56}
\end{split}
\end{equation}
as in Eq. \ref{eq:psi2pir} with 2 extra dimensions. This fixes the field to behave as
\begin{equation}\begin{split}
\Psi(x,y_5,y_6)&=e^{i\beta^a_5 y_5T_a/2\pi R_5+i\beta^a_6 y_6T_a/2\pi R_6}\ \tilde{\Psi}(x,y_5,y_6),\\ \tilde{\Psi}(x,y_5,y_6)&=\tilde{\Psi}(x,y_5+2\pi R_5,y_6)=
\tilde{\Psi}(x,y_5,y_6+2\pi R_6),
\end{split}\end{equation}
which is the equivalent of Eq. \ref{eq:solpsi} with 2 extra dimensions and commuting phases.
In the Poincar\'e algebra from Eq. \ref{eq:poin}, the translation operators commute
\begin{equation}
[P_5,P_6]=0.
\end{equation}
If we were to comply with these conditions when upgrading the momentum operators to include an element of the gauge algebra,
they would have to comply with Eq. \ref{eq:poinga}
\begin{equation}
\beta^a_5\beta^b_6f_{ab}^{\ \ c}=0,
\end{equation} 
which greatly diminishes the possible flavon alignments coming from Wilson lines. This equation is fulfilled when $\beta^a_5\propto\beta^b_6$ due to the antisymmetry of the lower indices, forcing both possible alignments to be equal. This equation is also fulfilled when the $\beta$'s are aligned in the directions where the structural constants vanish. Therefore we can have as many different alignments as the rank of the group. This is the usual Wilson line mechanism in arbitrary extra dimensions.

However the commutative requirement can be avoided. Non-commutative translations on EDs are well studied  \cite{Guralnik:2001pv}. This is usually done by choosing the EDs to be the so called non-commutative torus.
We could still have the usual (commutative) torus in extra dimensions, providing we introduce the generalized translation operators
whose commutator is
\begin{equation}
[P_5,P_6]=[\beta_5^a T_a,\beta_6^b T_b]=\beta^a_5\beta^b_6f_{ab}^{\ \ c} T_c=\theta_{56},
\end{equation}
where $\theta^{56}$ is the fundamental object for non commutative geometry. However we have a special case for $\theta^{56}$ since it is an element of the gauge algebra. We have commutative translations up to a phase.
We can have a usual commutative torus, but a non-commutative fibre bundle. The fields that live in the spacetime only, with trivial section on the fiber bundle (gauge singlets), only feel the usual translation operators of the usual commutative torus. The fields that are not gauge singlets obtain a phase which has a non-commutative dependence of the EDs. This implies that the fibre bundle is built from the usual 4d spacetime $\mathcal{M}^4$, the usual commutative torus $T^2$, and the gauge group $G$ through a non trivial multiplication of them $\mathcal{M}^4\times (T^2\ltimes G)$. This way we can have a generalization without having to comply with all the commutative relations for Wilson lines and orbifold boundary conditions \cite{deAnda:2018yfp}.

Consider the $\star$, the Moyal star product \cite{Guralnik:2001pv}
\begin{equation}
f(z)\star g(z)=e^{i\theta_{56}\partial_5\partial_{6'}}f(z)g(z')|_{z'\to z},
\end{equation}
where $z=(y_5,y_6)$ are the extra dimensional coordinates while $z$ and $z'$ are independent variables. This product allows us to find a consistent $\Psi$ from Eq. \ref{eq:psi56} with non-commutative phases as
\begin{equation}\begin{split}
\Psi(x,y_5,y_6)&=e^{i\beta_5^bT_by_5/2\pi R_5}\star e^{i\beta_6^bT_by_6/2\pi R_6}\tilde{\Psi}(x,y_5,y_6)\\
&=e^{i\beta_\star^b(y_5,y_6)T_b}\tilde{\Psi}(x,y_5,y_6)
\end{split}\end{equation}
where  $\tilde{\Psi}$ are normal commutative and periodic functions of the extra dimensions. This is just a phase $\beta_\star$ and can be reabsorbed through a gauge transformation 
\begin{equation}
\Psi(x,y_5,y_6)\to e^{-i\tilde{\beta}_*^b(y_5,y_6)T_b}\star\Psi(x,y_5,y_6)=\tilde{\Psi}(x,y_5,y_6),
\end{equation}
where $\tilde{\beta}_*^b$ is chosen to cancel the phase and it need not be equal to $\beta_*^b$ since the phases are mutiplied with the non trivial Moyal product. The gauge fields transform as
\begin{equation}
A_{5,6}\to \tilde{A}_{5,6}+(\partial_{5,6} e^{i\beta_\star^bT_b})\star e^{-i\tilde{\beta}_*^b(y_5,y_6)T_b},\end{equation}
which after compactification is just a constant and behaves like an effective 4d constant VEV
\begin{equation}
\braket{A_{5,6}}=\int dy^2 (\partial_{5,6} e^{i\beta_\star^bT_b})\star e^{-i\tilde{\beta}_*^b(y_5,y_6)T_b}
\end{equation}
as a consequence of the Wilson line mechanism. This way we have different alignments for the VEVs  $\braket{A_{5,6}}$, by choosing different noncommuting $\beta_{5,6}^b T_b$. This is the same Wilson line mechanism described in Eqs. \ref{eq:psi2pir}-\ref{Wilsonvev} replacing the product of noncommutative phases by the Moyal star product.

This is in line with the special case of noncommutative geometry, where the noncommutative tensor $\theta^{56}$ is just a gauge transformation. This is equivalent to the usual commutative geometry with non trivial boundary conditions in a non abelian group \cite{Saraikin:2000dn}. 

In general we assume that the compactification scale, identified with the flavour breaking scale, is quite high. After compactification the only trace of the noncomutativity we have is the noncommuting Wilson lines and only usual commuting functions with normal products. With this setup we can obtain effectively as many flavons as extra dimensions, where each flavon associated with each extra dimension has an independent vacuum alignment.

\section{Conclusion}
\label{conclusion}

In this paper we have proposed the idea that flavons can emerge from extra dimensional gauge fields, 
referred to as GFU analogous to GHU.
In such an approach there is a gauged family symmetry in extra dimensions and the flavons are the extra dimensional components of the gauge field. This identifies the compactification scale with the flavour breaking scale which is much larger than the electroweak scale, making such an idea viable in principle. 
In 5d models, this provides a simple mechanism to align the VEVs of the flavons through a combination of Wilson line and orbifold symmetry breaking, with the Poincar\'e group structure determining the relation of the different resulting flavon VEVs. 
No flavons need to be added by hand, nor is any alignment potential required.

We have presented some simple 5d examples of GFU based on $SO(3)$ and $SU(4)$ gauged family symmetry. 
In the simplest case of $SO(3)$, the adjoint representation is a triplet, leading a flavon triplet with non-trivial VEV alignment.
In the case of $SU(4)$ we have shown how this gauge group is broken upon compactification (through orbifolding 
and Wilson lines) to an $SU(3)\times U(1)$ gauge group in 4d,
together with a triplet and antitriplet flavons which develops a VEV, further breaking the rank of the gauge group. We have taken the first steps to constructing a model along these lines by introducing fermions and showing how they may couple to the flavons in a consistent way via heavy KK modes, leading to effective higher order operators
which play the role of Yukawa couplings.

Finally, we have presented a general formalism for Wilson lines and orbifolds, in any number of dimensions, including non-commutative
fibre bundles, which may be useful for aligning additional flavons as required for constructing realistic models.

\subsection*{Acknowledgements}
We would like to thank Patrick K.S. Vaudrevange for carefully reading the manuscript and for his useful comments.
SFK acknowledges the STFC Consolidated Grant ST/L000296/1
and the European Union's Horizon 2020 Research and Innovation programme under Marie Sk\l{}odowska-Curie grant agreements Elusives ITN No.\ 674896 and InvisiblesPlus RISE No.\ 690575. AA acknowledges support from CONACYT project CB-2015-01/257655 and SNI (México).

\end{document}